\documentclass[preprint,12pt]{elsarticle}

\usepackage{amssymb}
\usepackage{amsmath}
\usepackage{algorithm}
\usepackage{algpseudocode}
\usepackage{graphicx}
\usepackage{multirow}

\begin{document}

\begin{frontmatter}



\title{MACAW: A Causal Generative Model for Medical Imaging}


\author[1,2]{Vibujithan Vigneshwaran\corref{cor1}} 
\ead{vibujithan.vigneshwa@ucalgary.ca}
\author[1,2]{Erik Ohara} 
\author[1,2,3,4,5]{Matthias Wilms\fnref{fn1}} 
\author[1,2,5]{Nils Forkert\fnref{fn1}} 

\cortext[cor1]{Corresponding author}
\fntext[fn1]{Matthias Wilms and Nils Forkert share the last authorship}

\affiliation[1]{organization={Department of Radiology, University of Calgary},
            addressline={2500 University Dr NW}, 
            city={Calgary},
            postcode={T2N 1N4}, 
            state={Alberta},
            country={Canada}}

\affiliation[2]{organization={Hotchkiss Brain Institute, University of Calgary},
            addressline={2500 University Dr NW}, 
            city={Calgary},
            postcode={T2N 1N4}, 
            state={Alberta},
            country={Canada}}

\affiliation[3]{organization={Department of Pediatrics, University of Calgary},
            addressline={28 Oki Dr}, 
            city={Calgary},
            postcode={T2N 6A8}, 
            state={Alberta},
            country={Canada}}

\affiliation[4]{organization={Department of Community Health Sciences, University of Calgary},
            addressline={3330 Hospital Dr NW}, 
            city={Calgary},
            postcode={T2N 4Z5}, 
            state={Alberta},
            country={Canada}}

\affiliation[5]{organization={Alberta Children’s Hospital Research Institute, University of Calgary},
            addressline={28 Oki Dr}, 
            city={Calgary},
            postcode={T2N 6A8}, 
            state={Alberta},
            country={Canada}}

\begin{abstract}
Although deep learning techniques show promising results for many neuroimaging tasks in research settings, they have not yet found widespread use in clinical scenarios. One of the reasons for this problem is that many machine learning models only identify correlations between the input images and the outputs of interest, which can lead to many practical problems, such as encoding of uninformative biases and reduced explainability. Thus, recent research is exploring if integrating \textit{a priori} causal knowledge into deep learning models is a potential avenue to identify these problems. However, encoding causal reasoning and generating genuine counterfactuals necessitates computationally expensive invertible processes, thus restricting analyses to a small number of causal variables and rendering them infeasible for generating even 2D images. To overcome these limitations, this work introduces a new causal generative architecture named Masked Causal Flow (MACAW) for neuroimaging applications. Within this context, three main contributions are described. First, a novel approach that integrates complex causal structures into normalizing flows is proposed. Second, counterfactual prediction is performed to identify the changes in effect variables associated with a cause variable. Finally, an explicit Bayesian inference for classification is derived and implemented, providing an inherent uncertainty estimation. The feasibility of the proposed method was first evaluated using synthetic data and then using MRI brain data from more than 23000 participants of the UK biobank study. The evaluation results show that the proposed method can (1) accurately encode causal reasoning and generate counterfactuals highlighting the structural changes in the brain known to be associated with aging, (2) accurately predict a subject's age from a single 2D MRI slice, and (3) generate new samples assuming other values for subject-specific indicators such as age, sex, and body mass index. The code is available for a toy dataset at the following link: https://github.com/vibujithan/macaw-2D.git.
\end{abstract}







\end{frontmatter}



\section{Introduction}
\label{sec:Introduction}

Recent advances in medical imaging and the emerging availability of digital health records have resulted in an abundance of data in healthcare. This wealth of data, along with the continuing increase in computing resources, has helped the field of medical image analysis enter a new era — deep learning. This leap holds significant promise for disease prevention, diagnosis, and treatment planning \cite{maceachern_machine_2021}. However, the translation of deep learning techniques from academic research into clinical deployment has been slow \cite{winder2024challenges}. The primary challenge often lies in the mismatch between the data used for training and the data encountered in real clinical scenarios \cite{castro_causality_2020}. This mismatch often causes models that perform well in research settings to generalize poorly when applied to clinical environments. Furthermore, the limited clinical translation of deep learning models can be partly attributed to their ``black box" nature, which lacks inherent explainability for their decisions  \cite{Vercio_2020}. 

Both of these issues can be primarily related to the way these models are trained. More precisely, these discriminative deep learning techniques are designed to maximize accuracy on a given dataset. This approach encourages models to exploit all possible correlations in the data, including shortcuts, to improve performance on that specific dataset. As a result, these models often capture spurious correlations and uninformative biases present in imaging data \cite{Souza2024bias}. Thus, while they excel in research settings, they frequently fail to generalize effectively or provide meaningful explanations for their predictions when they are deployed in clinical settings.

One approach to better analyze and interpret the available medical data is to move beyond correlation-based analyses and instead utilize causally informed models. This approach has not only the potential to identify which variables (\textit{e.g.,} age, sex, race, scanner type, \textit{etc.}) causally affect medical images but also to what extent. Within this context, causal discovery methods address the question of \textit{which} variables have an impact, while causal reasoning methods explore \textit{how} they affect the images \cite{sanchez_causal_2022}. While causal discovery is beyond the scope of this work, we demonstrate in this work how true causal reasoning can be achieved using medical images through generative modeling. This causal generative modeling provides a powerful framework to explore variables related to the data-generating process in a transparent manner. 

A straightforward approach to causal generative modelling is integrating a known structural causal model (SCM) into an existing generative framework. Pearl \cite{pearl_causal_2012} describes a causal model as comprising three levels of complexity, referred to as the ``causal ladder". These levels, in order of increasing complexity, are association, intervention, and counterfactuals. \textit{Association} pertains to correlations in the data and is solely focused on modeling the probability distribution in the dataset. \textit{Intervention} relies on structural assumptions about the underlying data-generation process and involves exploring interactions with variables to observe how outcomes change on a population level. Many existing deep generative models, such as conditional VAEs, conditional GANs, and causal generative neural networks, only fulfill the requirements up to the intervention level. Lastly, \textit{Counterfactuals} can be used to investigate hypothetical scenarios on an individual level, encompassing both interventional and associational inquiries. A counterfactual query essentially asks the trained causal generative model, ``How would a subject's data appear if it had been acquired under different conditions?". The generation of counterfactuals not only provides causal insights into metadata and medical images but also holds significant potential in tasks with practical applications such as fairness, bias mitigation, data augmentation, data harmonization, and digital twins. \cite{pawlowski_deep_2020}. 

While the term ``counterfactual" is frequently employed in literature, only a few models, such as Deep Structural Causal Models (DSCMs) \cite{pawlowski_deep_2020}, Neural Causal Models (NCMs) \cite{xia2022causal}, Hierarchical Variational Autoencoders (HVAE) \cite{ribeiro_high_2023}, VQ-VAE and generalized linear models \cite{peng2024latent3dbrainmri}, and Diffusion SCM (Diff-SCM) \cite{sanchez2022diffusioncausalmodelscounterfactual}, have the ability to produce causally-grounded counterfactuals by following the \textit{Abduction-Action-Prediction} steps as defined by Pearl \cite{pearl_causal_2012}. Other methods in this domain mostly focus on generating realistic adversarial images aimed at deceiving classifiers. This scarcity of true counterfactual models for images arises from the requirement for invertible deep networks to achieve genuine counterfactual generation. To date, only the normalizing flow model is invertible by nature and has been previously used for causal modeling \cite{pawlowski_deep_2020, ribeiro_high_2023}. In this series of studies, each causal variable within a graph is modeled by a separate conditional normalizing flow, necessitating multiple normalizing flow models to represent the complete causal framework.

Alternatively, our work demonstrates, for the first time, that a single normalizing flow model coupled with masked autoencoders is sufficient to efficiently model complex causal structures.  While we only consider three causal variables in the experiments in this work, the approach presented can be easily extended to any number of causal variables by defining the corresponding adjacency matrix. Within this context, we demonstrate that combining a standard dimensionality reduction technique, such as kernel principal component analysis (KPCA), with normalizing flows can effectively encode a given causal structure. In theory, the KPCA technique can be replaced with any dimensionality reduction technique. Within its medical context, relevant previous studies are the works by \cite{wilms_invertible_2022} and \cite{bannister_deep_2022} on counterfactual image generation and Bayesian classification. The key difference to the work presented here lies in their exclusive focus on a setup assuming independence between conditioning variables, while the method presented in this work encodes a complex causal structure of interactions between variables using masked autoencoders. Thus, the main contributions of this work can be summarized as follows:

\begin{enumerate}
\item We present and evaluate a novel method named masked causal flow (MACAW) for encoding the causal structures of the data-generating process into a generative model.

\item Using this model, we generate and evaluate high-resolution counterfactuals associated with brain aging.

\item We also generate new high-resolution brain images created through causal interventions.

\item The model's explicit density estimation enables direct Bayesian classification, eliminating the need for a separate discriminative model.

\end{enumerate}

\section{Background}

\subsection{Causal graphical models}

Causal graphical models employ nodes to depict variables and edges to illustrate their connections within a directed acyclic graph (DAG) to provide an intuitive method for defining and exploring dependencies. The two important conditions, the Markov condition and faithfulness condition, ensure that conditional independence in the joint probability distribution is accurately reflected in the causal graphical model. Readers interested in these concepts can find in-depth information in the book by \cite{peters_elements_2017}.

Consider the $d$-dimensional random vector $\textbf{x} = [x_1, \dots , x_d]^T \in \mathbb{R}^d$, which follows the distribution $p_\text{x}(\textbf{x})$. Let a causal graphical model $\mathcal{G}$ be a DAG containing $d$ nodes, each represented by $x_i \in [1,d]$; the connection between nodes is defined by adjacency matrix $A \in {\{0, 1\}}^{d\times d}$. It is essential to emphasize that the adjacency matrix of any topologically ordered DAG is always triangular. Assuming Markovianity, $\mathcal{G}$ is a valid representation of $p_\text{x}(\textbf{x})$ if and only if the probability density $p_\text{x}(\textbf{x})$ can be factorized as follows:

\begin{equation}\label{eq:graphical_model}
p_\text{x}(\textbf{x}) = \prod_{i=1}^{d} p_\text{x}(x_i|\pi(x_i))
\end{equation}

Here, $\pi(x_i)$ denotes the set of parents of the node $x_i$, where $\pi(x_i) =\{x_j; A_{j,i}=1\}$. The basic network structure (adjacency matrix $A$) is typically established by making use of existing causal knowledge in the field. In cases with no or limited existing causal knowledge about the data-generating process, causal discovery algorithms can be utilized for this purpose \cite{glymour_review_2019}. 

The typical method for modeling a probabilistic causal model is to use Structural Equation Modeling (SEM) with random noise. The structural equation for each variable $x_j$ is defined as $S_j: x_j=f_j(\pi(x_j),n_j)$, where $n_j$ represents the mutually independent exogenous noise variable of the noise distribution $p_\text{n}$. In this formulation, the observational distribution of variables $p_\text{x}(\textbf{x})$ can be conceptualized as being generated by sampling from a noise distribution $p_n$ and then applying a set of structural equations $S$ to the sampled values. This implies that the observed variables are influenced by both, the noise distribution and the causal relationships represented by the structural equations.

The causal calculus \cite{pearl_causal_2012} was created for the utilization of a causal model. Specifically, the $do()$ operator enables an intervention in the model. When applying the $do()$ operator, specific functions are replaced with a constant in an SEM. Similarly, in the corresponding DAG, the edges going into the target of intervention are removed, but the edges exiting the target are retained.

\subsection{Normalizing flows}

Normalizing flows are used to model complex probability distributions, denoted as $p_\text{x}$, by applying a sequence of transformations $\textbf{T} = \textbf{T}_1 \circ \cdots \circ \textbf{T}_k$  to a simple density prior $p_\text{z}$. Transformations in $\textbf{T}$ must be both invertible and differentiable to allow training of the model using the change of variables formula:

\begin{equation} \label{eq:likelihood}
    p_\text{x}(\textbf{x}) = p_\text{z}(\textbf{T}^{-1}(\textbf{x}))| \text{det} J_{\textbf{T}^{-1}}(\textbf{x})|
\end{equation}

Here, $\text{det} J_{\textbf{T}^{-1}}$ is the determinant of the Jacobian of the inverse transformations. Efficient model optimization hinges on the ease of computing this determinant. As a result, Jacobians of the determinants are often designed as triangular matrices, allowing for computation in $\mathcal{O}(n)$ time. Various techniques have been introduced in the literature to achieve this triangular structure, with common approaches including the use of coupling and autoregressive functions.

Of particular relevance to this work, we briefly describe autoregressive functions next. Autoregressive functions transform a variable $x_i$ using variables $x_1$ to $x_{i-1}$, which in turn constrains the Jacobian of the transformation to be lower triangular. This is similar to writing a multivariate density $p_\text{x}$ as a product of univariate conditional densities:

\begin{equation} \label{eq:autoregressive}
 p_\text{x}(\textbf{x}) = p_\text{x}(x_1) \prod_{i=2}^{d} p_\text{x}(x_i|x_{1:i-1})
\end{equation}

Following this step, the flow model is trained directly through maximum likelihood optimization using equation \eqref{eq:likelihood}. More detailed descriptions about normalizing flows are, for example, provided by \cite{kobyzev_normalizing_2021}.

\subsection{Related work}
\label{sec:related}

This subsection outlines the role of related works in shaping the development of the proposed MACAW model. Specifically, our research is built upon the foundations laid by Wehenkel et al. \cite{wehenkel_graphical_2021} in their work on graphical flows and Khemakhem et al. \cite{khemakhem_causal_2021} in their study on causal flows.

Wehenkel and Louppe \cite{wehenkel_graphical_2021} highlighted the similarity between equations \eqref{eq:graphical_model} and \eqref{eq:autoregressive}, and argued that autoregressive transformations can be interpreted as a method to model a causal network with a predetermined node ordering. Conversely, in the case of a specific DAG with a predefined causal relationship, the autoregressive conditioners can be selectively masked to incorporate the causal relationship into the model. Based on this new perspective, they proposed the graphical normalizing flow technique, a new invertible transformation with either a prescribed or a learnable graphical structure to inject domain knowledge into normalizing flows. In a similar work, Khemakhem et al. \cite{khemakhem_causal_2021} suggested that SEMs and autoregressive flows are similar and introduced a framework called causal autoregressive flow (CAREFL) for causal discovery. Furthermore, they showed that normalizing flows can generate effective counterfactual queries due to their invertible nature. 

Both of these studies primarily focused on identifying the causal structure or topology within a given dataset, operating with limited variables. Specifically, Khemkhem's counterfactual implementation only permitted coupling flows, accommodating two sets of independent variables. In contrast, the primary focus of our work is to develop a method for classification and counterfactual generation for higher-dimensional datasets, such as images, and to encode complex (non-autoregressive) causal structures into the flows. Consequently, a model is required that scales effectively for larger dimensions and allows the parallel execution of the flow. Therefore, we introduce a neural network called the causally masked autoencoder (C-MADE) in this work, which is inspired by the masked encoder developed by Germain et al.\cite{germain_made_2015}. This implementation requires only a single forward pass to compute all causal dependencies and their respective conditional likelihoods, making it a computationally efficient density estimator compared to existing alternatives. Subsequently, multiple C-MADEs are arranged in sequence to form the Masked Causal Flow (MACAW), akin to how Papamakarios et al.\cite{papamakarios_masked_2017} utilized stacked MADEs to create masked autoregressive flows. The following section provides a detailed explanation of this method.

\section{Methods}
\label{sec:methods}
\begin{figure*}[h]
    \centering
    \includegraphics[width=\textwidth]{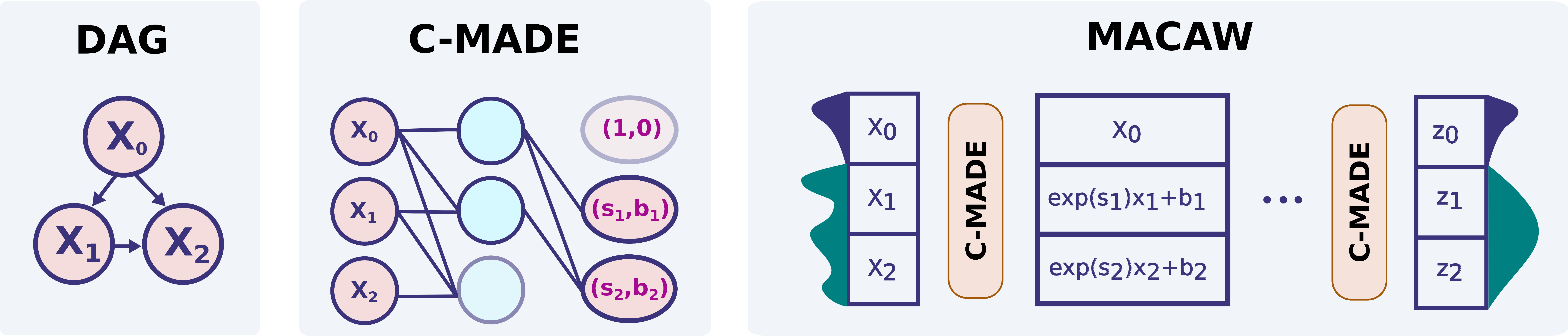}
    \caption[]{A causal DAG (left) and its
respective C-MADE network (center). The MACAW architecture (right) consists of multiple C-MADE networks connected in a series, thereby forming a normalizing flow.}
    \label{fig:macaw}
\end{figure*}

\subsection{Causally-masked autoencoders (C-MADE)}

We constructed C-MADE networks to efficiently represent the causal structure using a neural network for a given causal graphical model $\mathcal{G}$ with the adjacency matrix $A$. For simplicity, we assume that a neural network has the same number of input units ($\textbf{x}$) and output units ($\textbf{z}$). Since the output $z_i$ must depend only on its parents $\pi(z_i) \subset \textbf{x}$, there must be no computational path between the output unit $z_i$ and any non-parent input units. A convenient way of zeroing connections is multiplying each matrix element-wise with a binary mask matrix derived from $A$, whose entries that are set to zero corresponding to the connections that need to be removed. Thus, the masks are essentially responsible for satisfying the causal structure. 

Furthermore, we assume that our input is D-dimensional, \textit{i.e.,} $\textbf{x} = [x_1,\dots, x_D]$, where the adjacency matrix $A$ characterizes the causal graph. To impose the causal property, we first assign an integer to each unit in the input and output layers from 1 to D. After numbering all the units, causal constraints on each unit are simply imposed by masking the connections based on matrix $A$. 

While hidden layers can be added to this network, the causal parental structure has to be accurately maintained through the hidden layers. In this setup, $i^{th}$ hidden neuron in $j^{th}$ layer, $h_i^j$, is connected to its parents  $\pi(h_i^j) \subset \textbf{h}^{j-1}$ as well as $h_i^{j-1}$, which ensures that the learned hidden values are propagated to the final layer. This can be done by augmenting the adjacency matrix $A$'s diagonal units to 1 for all input-hidden layer connections and hidden-hidden layer connections. Moreover, the number of neurons in a hidden layer must be a multiple of $D$ to preserve the causal structure. If a hidden layer has $n \times D$ neurons, $A$ can be duplicated $n$ times and stacked to create the adjacency matrix for that layer. However, as the neurons in the final layer (output layer) are connected only to their parents, the network becomes strictly triangular, thus preserving the initial causal DAG structure. Figure \ref{fig:macaw} shows a causal graph and its corresponding C-MADE network.


\subsection{Masked Causal Flow (MACAW)}
Let us consider a normalizing flow model whose transformed probability density $p(\textbf{z})$ is given as the product of prior distributions. The prior distributions for the source variables (variables that do not have parents) should be selected to resemble their original distributions closely. This is because the distributions of these variables are not modified throughout the training process. A standard normal prior is assigned to all variables that are not sources. Thus, a non-source variable $z_i$'s distribution is given by $p(z_i | \pi(z_i)) = \mathcal{N}(0,1)$. This can be achieved by using the change of variable formula \eqref{eq:likelihood} with a series of affine transformations $T_t$, where $t \in [1, K]$. The transformed variables are denoted as $\textbf{m}_t$. For notation consistency, we assign $\textbf{x} = \textbf{m}_0$ and $\textbf{z} = \textbf{m}_K$. Each affine transformation $T_t$ has scale and shift vectors, $\textbf{s}_t$ and $\textbf{b}_t$, estimated by a unique C-MADE network $C_t$. 

\begin{equation} \label{eq:cmade}
(\textbf{s}_t , \textbf{b}_t) = C_t(\textbf{m}_{t-1})
\end{equation}

\begin{equation} \label{eq:scale}
 \textbf{m}_{t} = exp(\textbf{s}_t)\textbf{m}_{t-1} + \textbf{b}_t 
\end{equation}

Since C-MADE is constructed as a causal network, the predicted affine transformation parameters of a variable can be considered as a function of its parents. Thus, the likelihood of the transformed probability density complies with equation \eqref{eq:likelihood}. The negative log-likelihood of $p(\textbf{z})$ was used as the loss function to optimize the C-MADE networks' weights. Following the optimization, the likelihood of a data point can readily be measured using a forward flow. Stacking multiple C-MADEs in a sequence provides the flexibility required to transform a complex distribution to the defined distribution \cite{papamakarios_masked_2017}.

\subsubsection{Generative sampling}
\label{sec:sampling}
To generate data using the trained model, we start by sampling the source variables from their prior distribution. Next, we proceed to sample each of the other variables iteratively by performing a backward flow. Moreover, in the case of interventional sampling, we have the option to set a particular value for a variable and continue sampling the remaining variables as usual.

\subsubsection{Counterfactual inference}
\label{sec:counterfactual}
Counterfactual queries aim to assess statements about hypothetical situations of already existing observations. Let's assume that the MACAW was used to model the density of dataset $\text{X}$ containing data vectors [$\textbf{x}^0, \dots, \textbf{x}^N]$. Within this context, one observed data vector from the dataset is denoted as $\textbf{x}^{obs} = [x_1,\dots, x_D]$. For instance, if variable $x_j$ had taken the value $x_j = \alpha$ in our observed feature vector $\textbf{x}^{obs}$, counterfactual queries can be used to determine what the value of variable $x_i$ would have been. This is denoted as $ x_{i, x_j\leftarrow \alpha}$. According to Pearl \cite{pearl_causal_2012}, generating causal counterfactuals requires three steps: \textit{abduction}, \textit{action}, and \textit{prediction}. The \textit{abduction} step evaluates the probability distribution over latent variables $\textbf{z}^{obs}$ given observations $\textbf{x}^{obs}$. In our model, this can be simply done using a forward flow, computing the transformation $\textbf{z}^{obs} = \textbf{T}(\textbf{x}^{obs})$. The next step (\textit{action}) is to intervene and fix the value of $x_j$ to a specific value $\alpha$, denoted as $do(x_j=\alpha)$, which makes it independent of its causes $\pi(x_j)$. In this step, the corresponding value of change of $ x_{i, x_j\leftarrow \alpha}$ is adjusted in the transformed space $\textbf{z}^{obs}_j$. The final step (\textit{prediction}) is performed by computing an inverse transformation pass of the intervened $\textbf{z}^{obs}$ to generate the counterfactual feature vector $\textbf{x}^{cf}$, which is done through a backward flow. The algorithm \ref{alg:counterfactual_query} outlines the steps for the counterfactual process \cite{khemakhem_causal_2021}.

\begin{algorithm}
\caption{Counterfactual query }\label{alg:counterfactual_query}
\begin{algorithmic}
\Require observed data $\textbf{x}^{obs}$, cf variable $x_j$, and cf value $\alpha$
\State 1. Abduction - forward flow
\State $\textbf{z}^{obs} \gets T(\textbf{x}^{obs})$
\State 2. Action - change hypothetical values in the \textbf{z} space
\State(a) $z_{i,x_j\leftarrow\alpha}^{obs} \gets 
T(\alpha,x_{\pi(j)}^{obs})$
\State(b) $z_{i,x_j\leftarrow\alpha}^{obs} \gets z_i^{obs} \text{ for } i \neq j$
\State 3. Prediction - backward flow
\State \text{Return } $\textbf{x}_{x_j \leftarrow \alpha} \gets T^{-1}(\textbf{z}^{obs}_{x_j \leftarrow \alpha})$
\end{algorithmic}
\end{algorithm}

\subsubsection{Bayesian classification}
\label{sec:Bayesian}
Let's denote our input distribution $p(\textbf{x})$ as a joint distribution of a set of features $\textbf{f}$ and a parent variable $c$ that we need to classify, which can take one value from $\{c^1,\dots, c^R\}$. Thus, $p(\textbf{x})$ can be written as $p(c, \textbf{f})$. When we set $c=c^i$, the forward flow of the network provides the posterior of $p(c=c^i,\textbf{f})$. This can be easily computed for all possible values of c $\{c^1,\dots, c^R\}$ simply by setting the class variable accordingly. Thus, for a given class $c=c^i$, the posterior can be computed using Bayes' theorem as follows:
\begin{equation}
p(c=c^i | \textbf{f}) = \frac{p(c=c^i, \textbf{f})}{\sum_{c}p(c=c^j, \textbf{f})}
\end{equation} 
Then, the class with the maximum a-posteriori (MAP) is chosen as the predicted class label. It is important to note that for each class variable, it is necessary to evaluate all possible posterior values and perform a forward pass with each of them.

\subsection{Dimensionality reduction}
\label{sec:dimensionality}
As described earlier, normalizing flows are known for their resource-hungry nature. Consequently, to maintain the dimensions of the variables throughout the flows, a naive implementation of MACAW results in severe computational challenges. Therefore, when dealing with images, operating with all image pixels is impractical and even 2D images have to be projected onto a low-dimensional latent space before applying the MACAW model for density estimation. In this work, we utilized Kernel Principal Components Analysis (KPCA) to reduce the dimensionality of the training images. However, this technique can easily be replaced with any dimensionality reduction technique. The basic idea of this method is to project data that is not linearly separable onto a higher-dimensional space, thereby transforming it into a linearly separable form. We selected this technique for two primary reasons: (1) projected features are not linearly correlated with each other, as they are projections onto an orthogonal basis. This property reduces the dependence between projected image features, leading us to assume that this makes the optimization faster. (2) The inverse (preimage) of the projected KPCA features (here: polynomial kernel with a degree of 3) can be computed efficiently using existing algorithms.

\subsection{Quantitative evaluation}
\label{sec:evaluation}
Assessing the effectiveness of counterfactual image generation techniques poses challenges due to the absence of real-world ground truth. Monteiro et al. \cite{monteiro2023measuringaxiomaticsoundnesscounterfactual} and then Melistas et al. \cite{melistas_benchmarking_2024} proposed a framework for evaluating counterfactual images generated by various generative techniques. Their framework uses structural causal models and employs the counterfactual inference based on Pearl's \textit{Abduction-Action-Prediction} steps to assess the performance of the different methods. In line with their study, we also conduct a specific set of experiments to evaluate the quality of our counterfactuals in terms of \textit{Realism} and \textit{Effectiveness}.

\subsubsection{Realism}
We used the Fréchet Inception Distance (FID) to quantify the semantic similarity between the generated counterfactual images and images in the training set. Therefore, real and generated samples were passed through an Inception v3 model \cite{szegedy_rethinking_2015} (pre-trained on Imagenet) to extract their semantic features and to calculate the FID between these two feature representations. A lower FID indicates that the features contain similar semantic information.

\subsubsection{Effectiveness}
Effectiveness aims to assess how well a counterfactual query performs. To quantitatively evaluate the effectiveness of a particular counterfactual image, we trained a separate discriminative deep learning model on the training set to predict the value of the intervened variable based on the image. While this traditional inference model may capture spurious correlations in the training data, it still provides information about the degree of confidence in the counterfactual generations.

\section{Experiments and Results}
\label{Experiments}

We performed two experiments to investigate the effectiveness of MACAW. In the first experiment, we used synthetic data with a known causal structure to demonstrate that likelihood estimation, intervention, and counterfactual analysis function as expected. In the second experiment, we utilized the UK Biobank brain MRI images to showcase the practical applications and benefits of interventional sampling, counterfactual inference, and Bayesian classification that are only possible with a true causal deep learning framework.

\subsection{Synthetic data}

\subsubsection{Data}
The first experiment aimed to evaluate if the proposed MACAW model can effectively learn the given causal structure within a dataset and accurately perform interventional and counterfactual queries. Therefore, we created a synthetic dataset with predefined structural equations for sample generation to investigate this in detail. The causal structure of this dataset and the specific structural equations are defined as follows:

\begin{equation}
\label{eq:scm}
\vcenter{\hbox{\includegraphics[height=2.5cm]{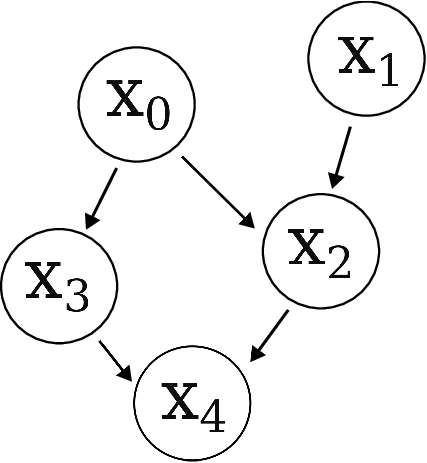}}}
\qquad\qquad
\begin{aligned}
x_0 &= n_0 
\\
x_1 &= n_1
\\
x_2 &= 2x_0 + x_1 + n_2
\\
x_3 &= 2x_0 + n_3
\\
x_4 &= 6x_2x_3 + n_4   
\end{aligned}
\end{equation}

Where the noise variable $n_i$ was sampled from uniform $\mathcal{U}$ and normal $\mathcal{N}$ distributions as follows:

\begin{equation}
\begin{aligned}
n_0 &\sim \mathcal{U}(0,1) & n_1 &\sim \mathcal{N}(1,1) & n_2 &\sim \mathcal{N}(0,2) \\
n_3 &\sim \mathcal{N}(0,0.5) & n_4 &\sim \mathcal{N}(0,0.1)
\end{aligned}
\end{equation}

We generated 10,000 samples using equation \ref{eq:scm} and partitioned it for training (70\%) and testing (30\%). Next, we defined prior distributions for the latent variables as follows: $z_0$ - drawn from a uniform distribution $\mathcal{U}(0,1)$, $z_1$ - drawn from a normal distribution $\mathcal{N}(1,1)$, while all other latent variables followed standard Gaussian distributions.

\subsubsection{Training}
For this experiment, our MACAW model consisted of 10 C-MADEs, each consisting of three hidden layers with 15 neurons each. During the training process, we utilized the negative log-likelihood as the loss function and halted training when reaching the point of minimal validation loss. 

\subsubsection{Generative sampling}

After model training, we generated 10,000 random samples using the MACAW model. The distribution from SEMs and those generated using MACAW were very similar. Table 1 outlines the mean and the variance of the variables generated.

\begin{table*}[h]
\label{tab:generated}
\caption{Mean and the variance of generated samples using SEM and MACAW.}
\centering
\begin{tabular}{|l|ll|ll|ll|}
\hline
\multirow{2}{*}{} & \multicolumn{2}{l|}{$x_2$}         & \multicolumn{2}{l|}{$x_3$}         & \multicolumn{2}{l|}{$x_4$}               \\ \cline{2-7} 
                  & \multicolumn{1}{l|}{SEM}   & MACAW  & \multicolumn{1}{l|}{SEM}   & MACAW  & \multicolumn{1}{l|}{SEM}      & MACAW     \\ \hline
Mean              & \multicolumn{1}{l|}{4.01} & 3.99 & \multicolumn{1}{l|}{2.99} & 2.99 & \multicolumn{1}{l|}{73.96}   & 74.55   \\ \hline
Variance          & \multicolumn{1}{l|}{5.32} & 5.92 & \multicolumn{1}{l|}{0.58} & 0.53 & \multicolumn{1}{l|}{2429.86} & 2863.50 \\ \hline
\end{tabular}
\end{table*}


\begin{figure}[ht]
    \centering
    \includegraphics[width=\textwidth]{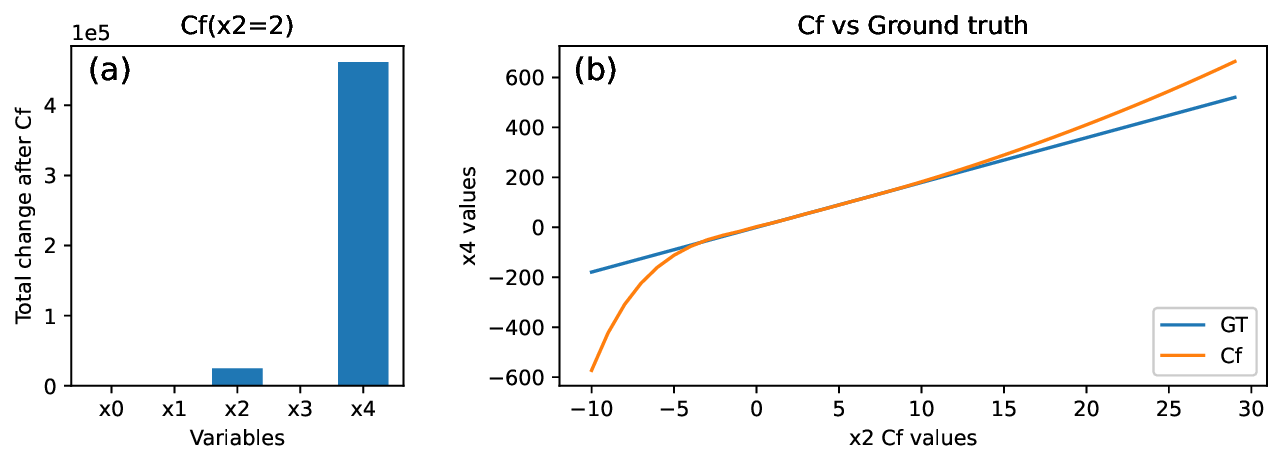}
    \caption[]{(a) The sum of the absolute difference between each variable in the test set after the counterfactual query for the action $do(x_2=2)$, (b) The expected values of the predicted counterfactual values and respective ground truth values.}
    \label{fig:exp1-cf}
\end{figure}

\subsubsection{Counterfactuals}
To measure the accuracy of the counterfactual inference quantitatively, we estimated the counterfactual values for the action $do(x_2=2)$ in the test set. We computed the absolute difference between the observed values and the counterfactual values and then summed them up. Fig. \ref{fig:exp1-cf}(a) shows that the counterfactual query only affected $x_2$ and $x_4$ while other variables remained unaffected, as expected based on the causal structure depicted in equation \ref{eq:scm}. Next, we compared the counterfactual values with their respective ground truth value. These ground truth values were directly determined by substituting $x_2=2$ in the equation $x_4 = 6x_2x_3^{obs} + n_4$, where the noise term $n_4$ was computed as $x_4^{obs} - 6x_2^{obs}x_3^{obs}$. Fig. \ref{fig:exp1-cf}(b) shows that the expected values of the ground truth and counterfactual values are perfectly aligned in the range of [-5,15], which is due to 99\% of $x_2$ values falling within this range during training.

\subsection{UK Biobank images}

\subsubsection{Data}
The second and main experiment used neuroimaging data from the UK Biobank (UKBB) cohort \cite{miller_multimodal_2016}. This study retrieved data under application 77508: Explainable and interpretable machine learning solutions in computational medicine. The UKBB's T1-weighted structural magnetic resonance imaging (MRI) used a 3D MPRAGE sequence with a 1-mm isotropic resolution and $208 \times 256 \times 256$ mm field of view. In the first step, starting with all subjects with available T1-weighted MRI data, participants with diagnosed brain-related disorders based on ICD10 codes (data field 41202, chapter V - Mental and behavioral disorders and chapter VI -Diseases of the nervous system) were excluded. We obtained the participants' sex information from the genetic sex data field (22001) and their age from the recorded values during the imaging visit (data field: 21003-2.0). Participants younger than 46 and older than 81 were excluded from the analysis because there was not a sufficient number of participants in these specific age groups to perform the age-stratified train-test split. Additionally, the body mass index (BMI) values were retrieved directly using the UKBB data field 21001. Subjects with NaN values for either sex, age, or BMI were excluded from this work, resulting in a sample size of 23,692 (male = 11,050, female = 12,642). 

All T1-weighted MRI data of the selected participants were aligned to the SRI24 atlas \cite{rohlfing_sri24_2010}, using the affine registration implemented in ANTs  \cite{avants_reproducible_2011}. From the 3D stack of data, we specifically chose the axial slice of a subject’s image that primarily covered the lateral ventricular region, an area of the brain that captures the extent of visible atrophy due to aging. Finally, the data was split into training and testing sets of 80\% and 20\%, stratified by age.

\subsubsection{Training}

\begin{figure}[ht]
\centering
\includegraphics[width=0.8\textwidth]{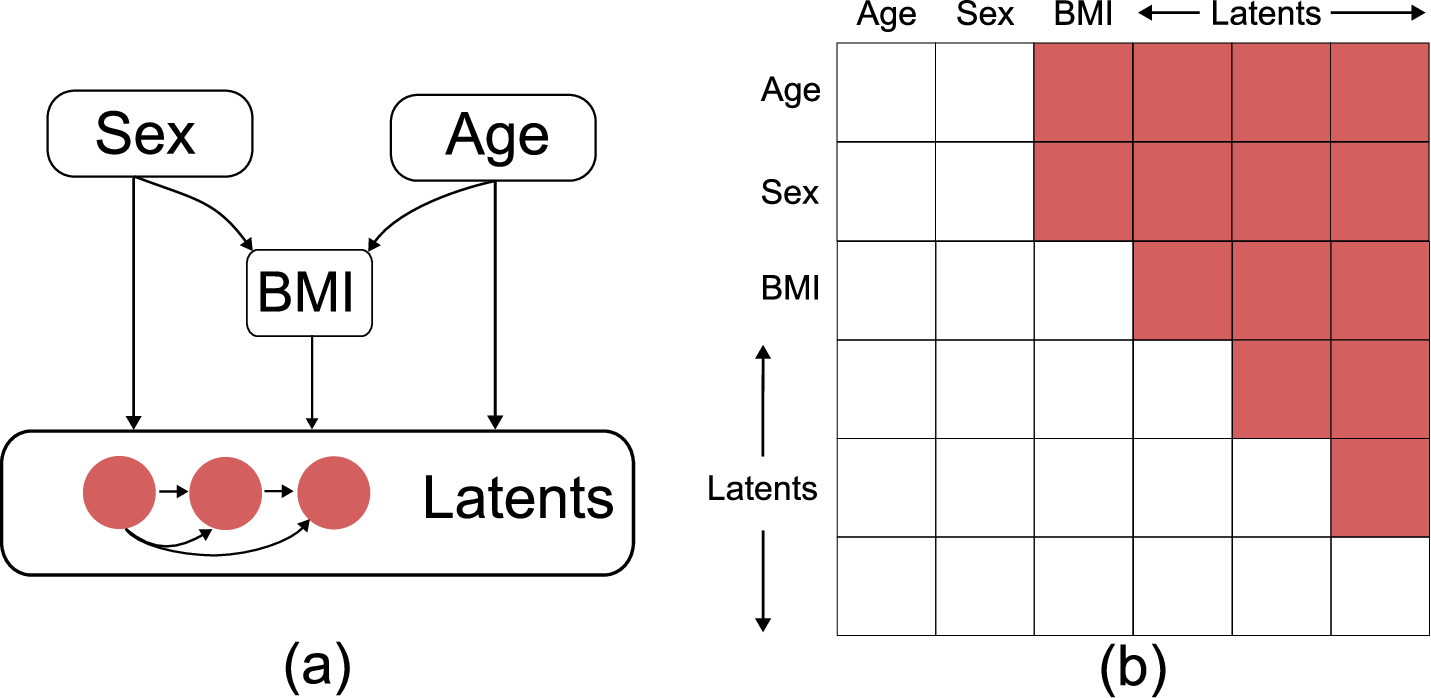}
\caption{(a) Predefined causal graph for the UKBB dataset, incorporating age, sex, and BMI values.
(b) Corresponding adjacency matrix, where filled cells indicate 1 and unfilled cells indicate 0.}
\label{fig:exp3-causal-graph}
\end{figure}

For model training, the extracted 2D images underwent center-cropping, leading to image sizes of $180 \times 180$ pixels. Subsequently, these images were projected onto a 1500-dimensional subspace (latents) using KPCA. Training all 1500 latents using the same MACAW network did not converge successfully, whereas training with 60 latents yielded a better likelihood estimation. Consequently, we divided the latents into subgroups of 60 and trained a separate MACAW model for each group, optimizing the likelihood individually for each model. Each MACAW model incorporated the causal structure illustrated in Figure \ref{fig:exp3-causal-graph}. In this setup, age and sex were treated as discrete distributions, while BMI was considered continuous. The prior distributions for sex and age were determined based on the distribution of the training data, adopting Bernoulli and categorical distributions, respectively. Gaussian distributions were used as priors for both BMI and the latents. A 10\% validation set was taken from the training set during the training phase to determine the best early stopping criteria.

\subsubsection{Generative sampling}
\begin{figure}[ht]
    \centering
    \includegraphics[width=0.7\textwidth]{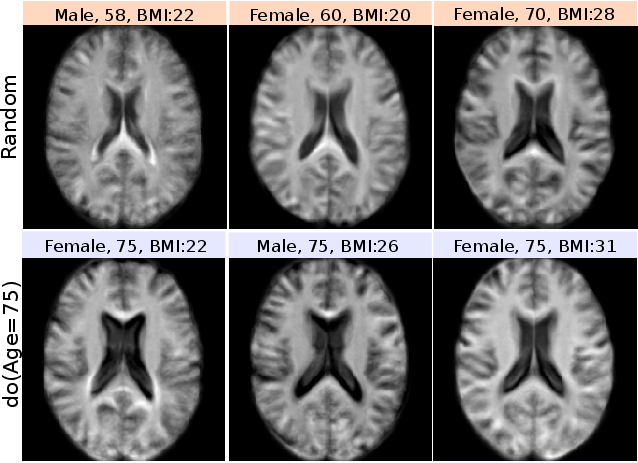}
    \caption[]{Results from two types of generative sampling: unconditional (top) and interventional (bottom). For interventional sampling, we set the age variable to 75 years.}
    \label{fig:exp3-sampling}
\end{figure}

Generative sampling was conducted in an autoregressive manner using the trained models. The process began by randomly sampling age, sex, BMI, and 60 components from the first model. This process continued iteratively until 1,500 components were generated, which were then reconstructed by estimating the preimage of the KPCA components. During interventional sampling, specific values for one or more parent variables (age, sex, BMI) were manually defined, while the remaining variables were sampled from the first model. This autoregressive sampling approach was then repeated using the subsequent models. The results from the generative and interventional sampling are presented in Figure \ref{fig:exp3-sampling}.

\subsubsection{Counterfactuals}
\begin{figure*}[ht]
    \centering
    \includegraphics[width=\textwidth]{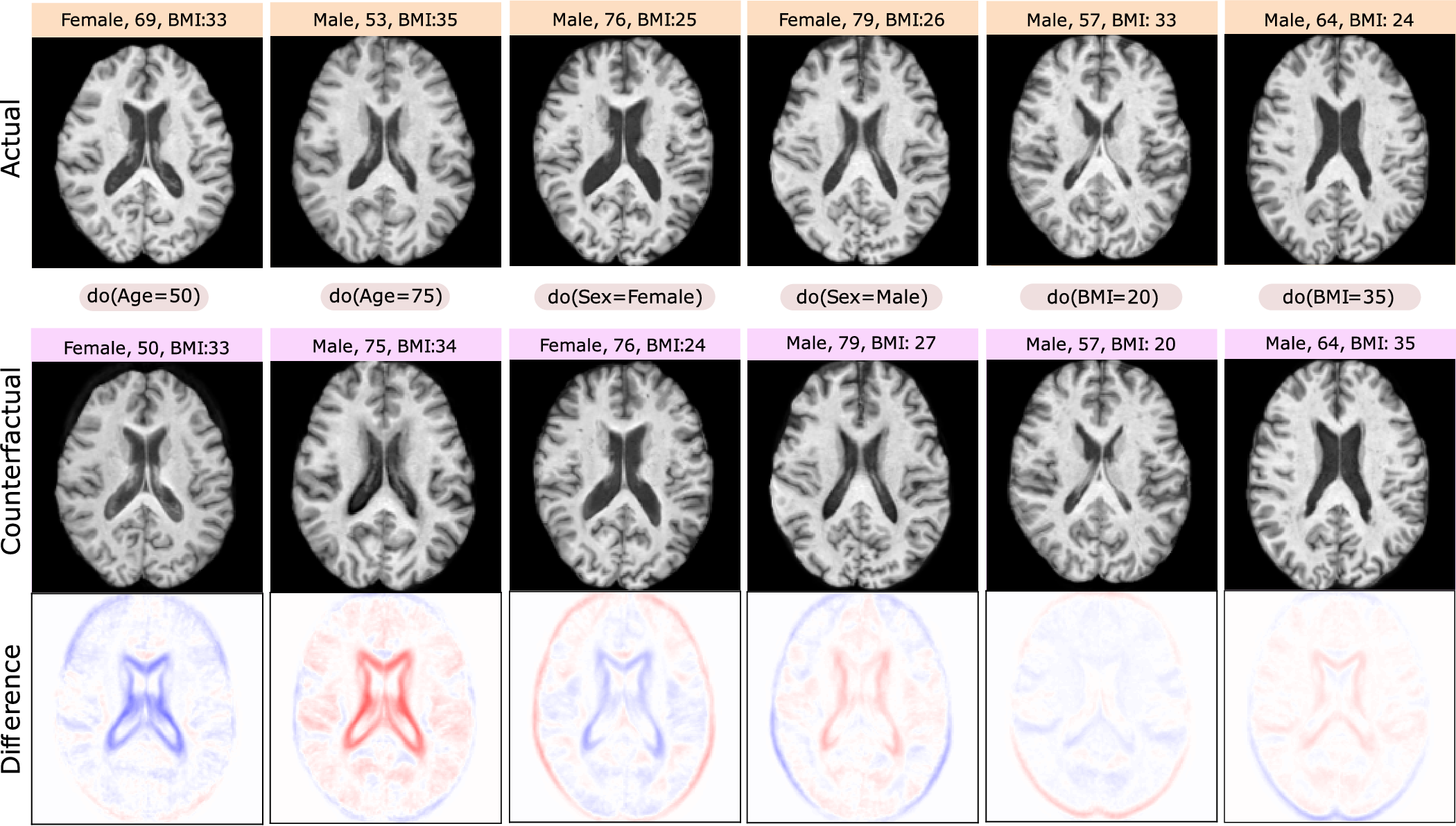}
    \caption[]{Illustration of the outcomes of counterfactual queries, where causal factors of randomly selected raw images were altered to generate new counterfactual images. Every column represents a distinct counterfactual query, identified by "$do(.)$" commands. }
    \label{fig:exp3-cf}
\end{figure*}

\begin{figure*}[ht]
    \centering
    \includegraphics[width=\textwidth]{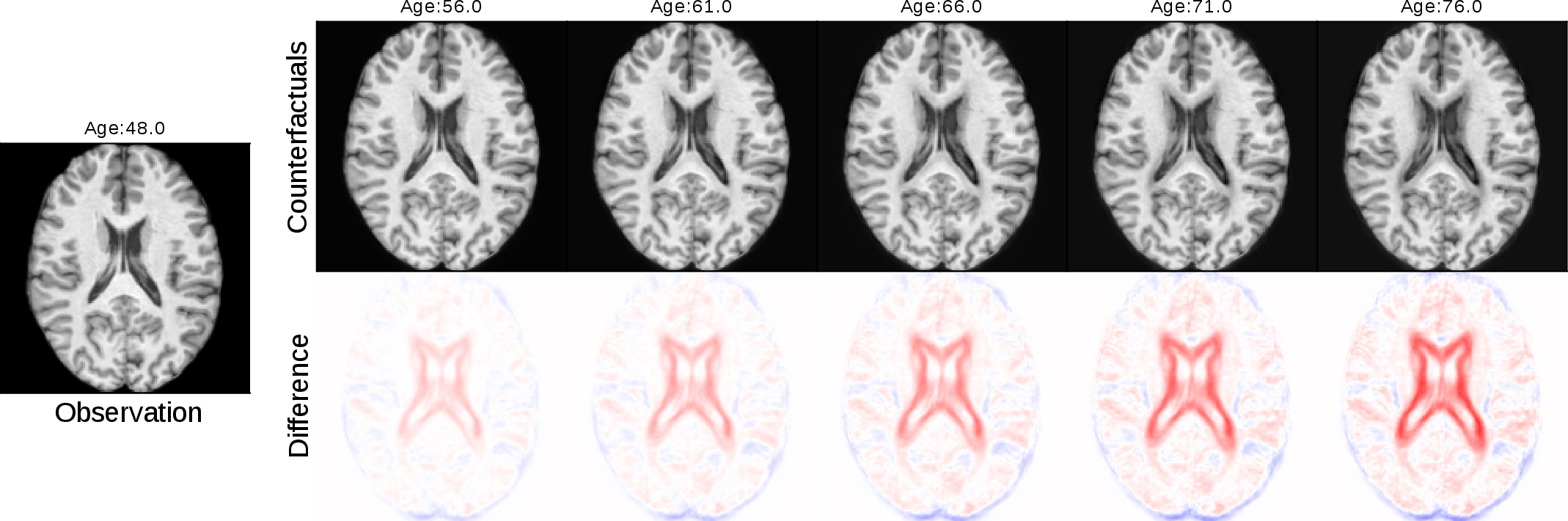}
    \caption[]{Results from counterfactual queries simulating brain aging. Each counterfactual image corresponds to a unique query ``$do(Age=x)$". }
    \label{fig:exp3-cf-progress}
\end{figure*}

Counterfactual inference was performed on data points within the test set. Using this setup, we aimed to estimate how an image would appear if the person had different biological characteristics, including age, sex, and BMI. Counterfactuals were generated independently for each model and then reconstructed, following a process similar to the generative sampling. Figure \ref{fig:exp3-cf} visually demonstrates that age has an expected impact on the ventricular and sulci regions. Increasing the age tends to enlarge the ventricular area while decreasing the age variable reduces the ventricular volume. Furthermore, changes in the sulci regions are also observed when changing the age variable. When considering sex-related counterfactuals, variations in brain size and subtle changes \cite{lotze_novel_2019} in the ventricular regions can be observed in the figure. Furthermore, when generating counterfactual images to illustrate sex differences, the BMI values appropriately change, with males exhibiting higher BMI values. This is attributed to the causal influence of sex on BMI. Finally, BMI changes result in alterations of the lateral and ventricular parts of the brain, which may be related to accelerated brain aging \cite{beck_cardiometabolic_2021}. In addition, Fig \ref{fig:exp3-cf-progress} displays results from counterfactual queries simulating brain aging.

To assess the effectiveness of counterfactual inference, we generated counterfactual age values for the entire test set, consisting of 4,739 images. Essentially, the query asks what would happen if everyone in the test set had age $\alpha$, denoted as $CF_{\alpha}$, where $ \alpha \in [55, 60, 65, 70, 75]$. The following experiments are conducted to assess these counterfactuals.
 
\textbf{Realism: }
In the first step, the FID between the training and test sets was measured to serve as the lower bound. Subsequently, all images in the test set were blurred using Gaussian blur with a standard deviation of 1, and the FID between each original test image and smoothed test image was measured as a comparative baseline. Finally, the distance between each $CF_{\alpha}$ defined above and the training set was calculated. The results displayed in Table 2 demonstrate that the generated counterfactual images show superior \textit{realism} compared to even slightly blurred images. Additionally, it was observed that \textit{realism} was notably high for the counterfactual images generated for the age of 65, which aligns with the centered age distribution of the data.

\begin{figure}[h!]
    \centering
    \includegraphics[width=\textwidth]{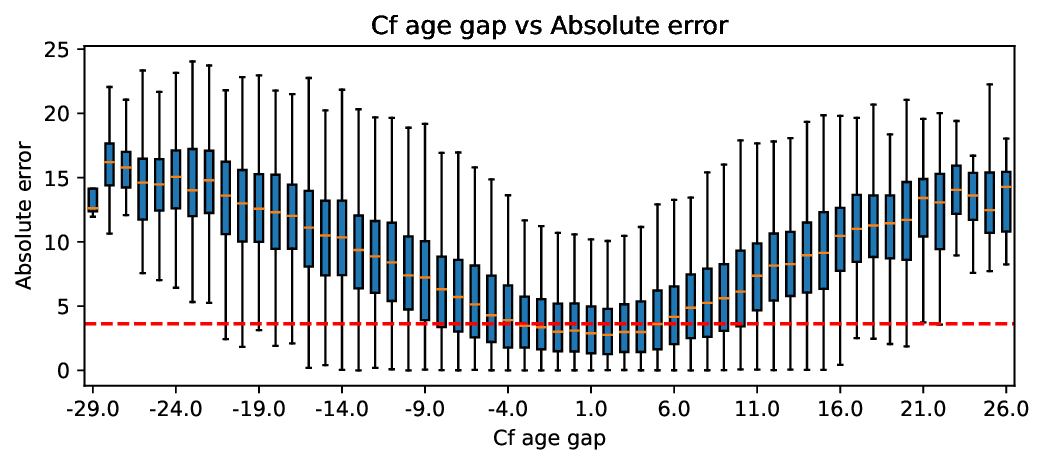}
    \caption[]{The figure displays how the effectiveness (MAE) of the generated images changes with the difference between actual age and queried counterfactual age. The dotted lines show the MAE of the test set (unaltered images).}
    \label{fig:cf_age_gap}
\end{figure}

\textbf{Effectiveness: }
To assess whether the generated counterfactuals indeed contain predictive age information, we trained a separate classifier trained on the entire training set. Specifically, we utilized a popular architecture for brain age estimation, the SFCN model \cite{peng_accurate_2021} trained for 2D slices, which achieved a mean absolute error (MAE) of 3.63 for test images. Subsequently, the $CF_{\alpha}$ sets were tested for age prediction using the counterfactual age $\alpha$ as the ground-truth target value for MAE computation. Table 2 displays these results, showing that $CF_{60}$ and $CF_{65}$ performed reasonably well. Since the UKBB has a large amount of training data for these age bins, it can be speculated that the model performs better when the queried counterfactual age is closer to the actual age (less severe changes on average). Therefore, to evaluate how the generated images change with counterfactual age differences, we computed the difference between the actual age and the counterfactual query age for all $CF_\alpha$ sets and subsequently measured the MAE value for this counterfactual age gap. Figure \ref{fig:cf_age_gap} illustrates these results, indicating that the counterfactual query performs well within the age range of -10 to +10.

\begin{table*}[]
\centering
\caption{ Realism and effectiveness are measured on test, blurred, and counterfactual images. FID: Fréchet Inception Distance, MAE: Mean Absolute Error. }
\label{tab:effectiveness}
\begin{tabular}{|l|l|l|l|l|l|l|l|}
\hline
 & Test & Blurred & $CF_{55}$ & $CF_{60}$ & $CF_{65}$ & $CF_{70}$ & $CF_{75}$ \\ \hline
Realism (FID) $\downarrow$ & 0.49 & 53.23 & 3.12 & 2.09 & 2.48 & 6.02 & 11.78 \\ \hline
Effectiveness (MAE) $\downarrow$ & 3.63 & - & 7.25 & 4.82 & 4.71 & 6.72 & 10.23 \\ \hline
\end{tabular}
\end{table*}

\subsubsection{Classification}

\begin{figure}[h!]
    \centering
    \includegraphics[width=0.9\columnwidth]{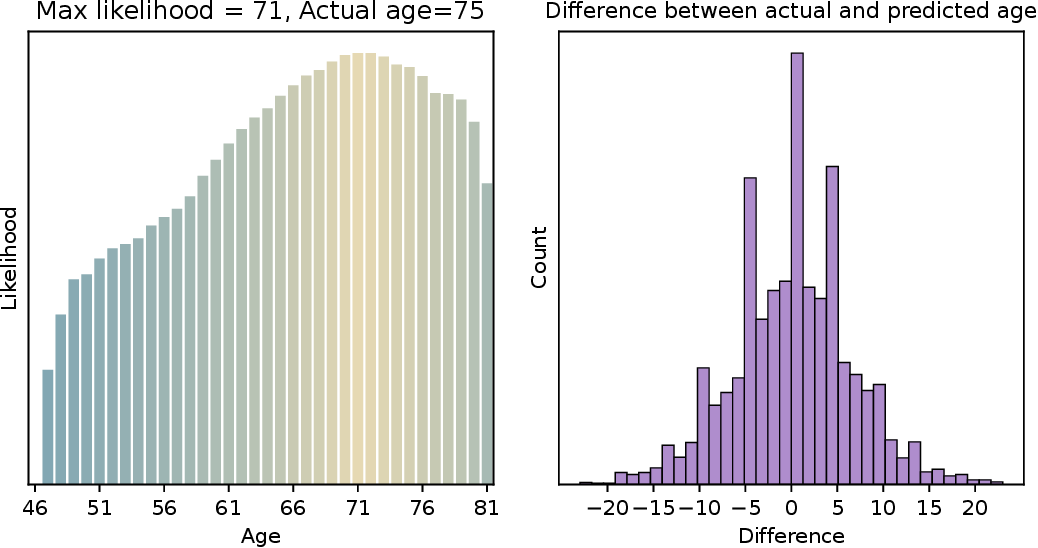}
    \caption[]{Posterior estimated by the model for each age, with the actual age being 75 and the predicted being 71 (left); distribution of the differences between the actual age and the predicted age for the whole test set (right).}
    \label{fig:exp3-class}
\end{figure}

Lastly, we employed the model's Bayesian classification abilities to predict the age of each participant in the test dataset as described in section \ref{sec:Bayesian}. Therefore, we computed the posterior for each age and selected the age with the maximum a-posteriori (MAP) value as the predicted age. The overall accuracy of this prediction resulted in an MAE of 5.047 (standard deviation (std) = 0.052) when we used the first model (60 latents) for classification. Figure \ref{fig:exp3-class} illustrates the posterior distribution for a participant and the disparity between actual and predicted values. The figures show that (1) the prediction assigns higher posterior values around the true chronological age and lower for distant ages, and (2) the error in the prediction distribution is centered around 0. These two results indicate that the model identifies the causal connection between age and images quite effectively.

\section{Discussion}
\label{discussion}

This work presents a novel causal generative framework called MACAW, which satisfies all three levels of the causal ladder (association, intervention, and counterfactual) and holds considerable promise for various applications in medical and clinical contexts. Specifically, counterfactual generation offers causal insights at the individual level, which cannot be achieved using correlation-based approaches. Counterfactuals on the MRI data revealed brain regions associated with aging, sex differences, and BMI changes that are all in line with current knowledge. 

Disentangling spurious correlations and identifying factors that causally influence medical images has been challenging. Discriminative deep learning models prioritize achieving accuracy on the training dataset, often exploiting shortcuts and spurious correlations. This makes it difficult to model and explore how variables like age, sex, and BMI causally affect medical images. In contrast, the causal generative modeling method described in this work encodes the data-generating process through a causal DAG, offering causal explanations at both the population level (interventions) and the individual level (counterfactuals). Furthermore, causal counterfactual generation offers theoretical benefits with respect to fairness \cite{kusner2018counterfactualfairness}, bias mitigation, data augmentation, data harmonization, and digital twin development. Furthermore, it is important to emphasize the distinction between conditional and causal approaches. In a standard conditional generation setup, the factors influencing image features are often assumed to be independent. For instance, when conditioning on the age variable, it is typically assumed that this has no impact on the BMI variable. However, in our setup, the model considers causal relationships within the dataset. As a result, when we intervene on the age variable, it correctly influences both, the BMI and the image features, reflecting genuine causal connections.

Thus far, researchers have used VAEs \cite{pawlowski_deep_2020, ribeiro_high_2023}, GANs \cite{nemirovsky_countergan_2021}, and diffusion model \cite{sanchez2022diffusioncausalmodelscounterfactual} for counterfactual image generation. However, for a causal model to truly perform counterfactuals, deterministic invertibility is essential. Normalizing flows offer this invertibility, and we demonstrated that a single normalizing flow combined with masked autoencoders is effective in modeling complex causal structures in our experiments. Quantitative assessments employing the metrics related to \textit{Realism} and \textit{Effectiveness} confirm that the generated counterfactual images effectively explain relevant causal information in the images. Furthermore, using interventional sampling, we created new samples with predefined parent values. This process can generate samples that align with the characteristics learned from the population. Consequently, these generated samples can serve as supplementary data for training other deep learning models, particularly in cases where certain classes have insufficient data. However, potentially more relevant for knowledge discovery in the biomedical context, it is possible to use the method proposed in this work as the basis for digital twins.

While the primary purpose of our model is counterfactual generation, it can also perform classification using Bayes' theorem, similar to other density estimation methods. However, its classification accuracy is lower compared to CNN-based methods like SFCN for brain age estimation \cite{peng_accurate_2021}. One potential reason for this finding is that, unlike task-specific models such as SFCN, our model's lower-dimensional subspace must capture features sufficient for accurate image reconstruction. However, the key advantage of our approach is that a single trained model can classify multiple variables (\textit{e.g.,} age, sex, BMI), whereas CNN-based methods require separate models for each variable. Although classification is not the model's primary focus, it may serve as a solid foundation for developing discriminative deep learning models in the future.

This study has several limitations, with one of the most notable being the reliance on a predefined causal graph. In our research, we assumed the availability of a causal graph for the UKBB dataset. However, a predefined causal graph may not be readily available in many real-world scenarios. Typically, researchers need to estimate causal relationships from existing domain knowledge or through randomized controlled trials, which may not always be feasible or ethical. In cases where a predefined causal graph is absent, causal discovery techniques can be employed \cite{glymour_review_2019}. Within this context, it is important to emphasize that estimating causal relationships solely from observational and cross-sectional data can be an extremely challenging task, often even impossible. Additionally, to capture the complex structure of the brain associated with aging, it is necessary to extend our method to 3D images, which still remains as our future work. Another potential limitation of the proposed model relates to the computational demands of the classification task. Specifically, for each class variable, it is necessary to evaluate all possible values and perform a forward pass with each of them. Consequently, this process incurs a computational cost several times greater (equal to the number of classes) than the typical computational load associated with a standard discriminative network. Moreover, this classification process is limited to anti-causal prediction, meaning it involves predicting a top-level parent variable based on its effects.

In conclusion, the experimental results demonstrate the potential and effectiveness of MACAW in generating interventional and counterfactual images, as well as performing Bayesian classification. Future research should prioritize the development of a technique that seamlessly integrates dimensionality reduction and the normalizing flow framework into a single model. This integration, in turn, would facilitate the efficient processing of 3D images, ensuring that the model effectively captures all crucial information. The proposed technique holds potential for exploring and identifying potential new biomarkers for various diseases. 

\section*{Acknowledgements}
This work was supported by the Canada Research Chairs program, the River Fund at Calgary Foundation, the Parkinson's Association of Alberta, and the NSERC Discovery Grant program. We would like to thank Anthony Winder for proofreading the paper.

 \bibliographystyle{elsarticle-num} 
 \bibliography{ref}






\end{document}